\documentclass[aps, prd, preprint, amsfonts, floatfix]{revtex4}

\usepackage{graphicx}
\usepackage{amsmath,amsfonts}
\usepackage{epsfig,color}

\begin{document}
\newcommand{\be}{\begin{eqnarray}}
\newcommand{\ee}{\end{eqnarray}}
\newcommand\del{\partial}
\newcommand\nn{\nonumber}
\newcommand{\Tr}{{\rm Tr}}
\newcommand{\Str}{{\rm Trg}}
\newcommand{\mat}{\left ( \begin{array}{cc}}
\newcommand{\emat}{\end{array} \right )}
\newcommand{\vect}{\left ( \begin{array}{c}}
\newcommand{\evect}{\end{array} \right )}
\newcommand{\tr}{{\rm Tr}}
\newcommand{\hm}{\hat m}
\newcommand{\ha}{\hat a}
\newcommand{\hz}{\hat z}
\newcommand{\hx}{\hat x}
\newcommand{\tm}{\tilde{m}}
\newcommand{\ta}{\tilde{a}}
\newcommand{\tz}{\tilde{z}}
\newcommand{\tx}{\tilde{x}}
\definecolor{red}{rgb}{1.00, 0.00, 0.00}
\newcommand{\rd}{\color{red}}
\definecolor{blue}{rgb}{0.00, 0.00, 1.00}
\definecolor{green}{rgb}{0.10, 1.00, .10}
\newcommand{\blu}{\color{blue}}
\newcommand{\green}{\color{green}}



\title{The Wilson Dirac Spectrum for QCD with Dynamical Quarks}
\author{K. Splittorff}
\affiliation{The Niels Bohr Institute and Discovery Center, Blegdamsvej 17, DK-2100, Copenhagen
  {\O}, Denmark} 
\author{J.J.M. Verbaarschot}
\affiliation{Department of Physics and Astronomy, SUNY, Stony Brook,
 New York 11794, USA}

\date   {\today}
\begin  {abstract}

All microscopic correlation functions of the
spectrum of the Hermitian Wilson Dirac operator  
with any number of flavors with equal masses
are computed. In particular, we give explicit results for
the spectral density in the physical case with    
two light quark flavors. The results include
the leading effect in the discretization error and are given  for 
fixed index of the Wilson Dirac operator. They have been 
obtained starting from chiral Lagrangians for the generating function of the
Dirac spectrum. Microscopic correlation functions of the real
eigenvalues of the Wilson Dirac operator are computed following the
same approach.

\end{abstract}
\maketitle

\section{Introduction}

The deep chiral limit of QCD with two quark flavors and the intimately
related nature of spontaneous chiral symmetry breaking is of direct
phenomenological interest. Also for studies beyond the standard model such 
as in QCD like theories with many light flavors or where the fermions 
are outside the fundamental representation, the deep chiral limit is
central.  
By a remarkable series of numerical and analytic developments it is
now possible to access the  
chiral limit by means of lattice QCD. The work presented in this paper 
is an attempt to facilitate the next step to the deep chiral limit by
offering an exact analytic understanding of the average behavior of
the small  eigenvalues of the Wilson Dirac operator at nonzero lattice
spacing, $a$. The behavior of these eigenvalues is  essential 
for chiral symmetry breaking \cite{BC,Heller,Aokiclassic} as well as for the
stability of lattice QCD computations \cite{Luscher}.
   
We consider the eigenvalue density of the Wilson
Dirac operator in
the microscopic scaling
limit \cite{SV,RMT,RMT1,RMT2} where  the product  of the
eigenvalues and the four-volume, $V$, as well as the product
$a^2 V$ are kept fixed. This part of spectrum is 
uniquely determined by \cite{DSV}
global symmetries, their breaking and the $\gamma_5$-Hermiticity of the Wilson Dirac operator
\be
D_W^\dagger = \gamma_5 D_W \gamma_5.
\label{g5Herm}
\ee
Because of this Hermiticity relation, the eigenvalues of the Hermitian Wilson Dirac operator,
\be
\label{D5def}
D_5\equiv\gamma_5(D_W+m),
\ee 
are real. In addition to correlations of these eigenvalues, we will also analyze the real eigenvalues
of $D_W$  in the microscopic limit.

In a recent letter \cite{DSV} and a longer follow-up \cite{ADSVprd} we have
shown how the {\sl quenched} microscopic Wilson Dirac spectrum can be
obtained from the chiral Lagrangian including order $a^2$-effects
for Wilson fermions. Although the supersymmetric method used in 
\cite{DSV,ADSVprd} can be applied to any number 
of flavors, the proliferation of terms makes the method only 
practical for use in the quenched case. Already for one dynamical 
flavor it becomes rather tedious to deal with analytically \cite{ADSVNf1}.
 
In this paper we follow a different path, the graded eigenvalue
method, that results in simple
expressions for any number of flavors with equal quark mass. It is also
possible to write down compact expressions for all spectral
correlation functions. 
The graded eigenvalue method is based on the observation that the order $a^2$ terms
in the (graded) chiral Lagrangian can be linearized at the expense of  extra 
Gaussian integrations. This results in compact expressions for microscopic Wilson Dirac spectra
for any number of flavors and all correlation functions.
The method was originally developed to describe transitions
between different universality classes of Random Matrix Theories
\cite{GuhrJMath,Alfaro-1994,GuhrAnn,GuhrComm}. The result obtained this way 
is an expression  in terms of diffusion in superspace where $a^2$
plays the role of time.

 All results in this paper are given for fixed index of the Wilson
 Dirac operator, defined for a given gauge field configuration by 
\be
\label{defIndex}
\nu = \sum_{k} {\rm sign} (\langle k|\gamma_5|k\rangle).
\ee
Here, $|k\rangle$ denotes the $k$'th eigenstate 
of the Wilson Dirac operator. The microscopic eigenvalue density for fixed 
$\nu$ gives detailed information on the effect of a nonzero lattice spacing 
on the would be topological zero modes at zero lattice spacing.

The study of the spectrum also casts new light \cite{DSV,ADSVprd} on 
the additional low energy constants of the chiral Lagrangian which is 
the backbone of Wilson chiral perturbation theory as developed in 
\cite{SharpeSingleton,RS,BRS,Aoki-spec,Aoki-pion-mass,bar-08,shindler-09} (reviews of 
effective field theory methods at finite lattice spacings can be found
in \cite{sharpe-nara,Golterman}). By a match of the two-flavor 
results presented 
in this paper to the microscopic spectrum of the Wilson Dirac operator 
on the lattice, the value of the low energy constants can be measured.
The spectrum of the Hermitian Wilson Dirac operator in the $p$-regime
of Wilson chiral Perturbation Theory has been discussed in
\cite{sharpe} and the results at next to leading order have been fitted
to lattice data in \cite{NS}.

The paper is organized as follows. 
Starting from a chiral Lagrangian for spectra of the Hermitian Wilson Dirac operator
at nonzero lattice spacing, we derive compact expressions for all spectral
correlation functions for any number of flavors.
 In the second part of this paper,  we obtain expressions for the distribution of the chiralities
over the real eigenvalues of the Wilson Dirac operator. 
Some technical details involving Efetov-Wegner terms are
discussed in Appendix A, and in Appendix B
 we give explicit expressions for
partition functions in terms of an integral over a diffusion kernel.

\section{Spectral Properties of  the Hermitian Wilson Dirac Operator}
\label{sec:universality}

The generating function for $p$-point spectral correlation functions
of the eigenvalues of the Hermitian Wilson Dirac operator for QCD with
$N_f$ dynamical 
quarks in the sector of gauge field configurations with index $\nu$ is given by
(recall that $D_5=\gamma_5(D_W+m)$)
\be
Z^\nu_{N_f+p|p} = \left \langle {\det}^{N_f} (D_5)
\prod_{k=1}^p\frac {\det (D_5 +z_k)}
{\det (D_5  +{z_k}'-i\epsilon_k \gamma_5)} \right \rangle.
\label{zgen}
\ee
The average is over gauge field configurations with index $\nu$ weighted by
the Yang-Mills action. For $p=0$ this is just the $N_f$ flavor partition function.
We will evaluate this generating function in the microscopic limit
where $V\to\infty$ with  
\be
mV, \quad z_kV,  \quad {z_k}'V, \quad a^2V   
\ee
kept fixed. 
The axial masses $z_k$ are required  when we apply the graded method
to obtain $p$-point  eigenvalue correlation functions
of the Hermitian Wilson
Dirac operator. For example,  from the graded generating function
${\cal Z}^\nu_{N_f+1|1}(m,z,z';a)$
we can obtain the spectral resolvent
\be
\label{Gsusy}
G^\nu_{N_f+1|1}(z,m;a)= \lim_{z'\to z} \frac{d}{dz} {\cal Z}^\nu_{N_f+1|1}(m,z,z';a),
\ee
and the density of eigenvalues, $\rho^\nu_5(\lambda^5,m;a)$,  of $D_5$ follows from 
\be
\rho^\nu_5(\lambda^5,m;a) = \left \langle \sum_k \delta(\lambda^5_k-\lambda^5) \right \rangle_{N_f} = \frac{1}{\pi}{\rm Im}[G^\nu_{N_f+1|1}(-\lambda^5)]_{\epsilon\to0}.
\label{rho5def}
 \ee
To derive expressions for the correlation functions 
in the microscopic limit we will rely on the graded eigenvalue method. 
 Before deriving the general result, we will first consider the case $p=0$, which is just the $N_f$-flavor partition function.

The chiral Lagrangian for Wilson chiral perturbation theory 
to $O(a^2)$ was
derived in \cite{SharpeSingleton,RS,BRS}. In \cite{DSV} we obtained 
the microscopic partition function
for fixed index $\nu$ also to order $a^2$, by decomposing the partition function
according to
\be
Z_{N_f}(m,\theta; a) \equiv \sum_{\nu=-\infty}^\infty e^{ i\nu\theta} Z_{N_f}^\nu(m;a). 
\ee
In the microscopic domain, the partition function reduces to a unitary matrix 
integral
\be
Z_{N_f}^\nu(m,z;a) =   \int_{U(N_f)} \hspace{-1mm} d U \ {\det}^\nu U
~e^{S[U]}, \label{Zfull}
\ee
where the action $S[U]$ for degenerate quark masses is given by
\be\label{lfull}
S & = & \frac{m}{2}\Sigma V{\rm Tr}(U+U^\dagger)+
\frac{z}{2}\Sigma V{\rm Tr}(U-U^\dagger)\\
&&-a^2VW_6[{\rm Tr}\left(U+U^\dagger\right)]^2
     -a^2VW_7[{\rm Tr}\left(U-U^\dagger\right)]^2 
-a^2 V W_8{\rm Tr}(U^2+{U^\dagger}^2) .\nn
\ee
 In addition to the chiral condensate, $\Sigma$, the action also contains the low
 energy constants $W_6$, $W_7$ and $W_8$ as parameters.
Since the $W_6$ and $W_7$ terms can be eliminated at the expense of an extra 
integration \cite{ADSVprd}, we will only consider the $W_8$ term
in the remainder of this paper. 
  For the reasons discussed in section VII of \cite{ADSVprd} we consider
  only $W_8>0$.
To simplify our notation below, we will absorb the factor $V
W_8$ into $a^2$  and the factor $V\Sigma $ into $m$, $z_k$ and $\lambda^5$ 
\be
  a^2 V W_8\to a^2, \qquad   mV \Sigma \to m,  \qquad
 z_k V \Sigma \to z_k\quad {\rm and} \quad \lambda^5 V\Sigma \to \lambda^5.
\ee

Up to a normalization factor, the term proportional to $W_8$ in the action (\ref{lfull}) can be rewritten as
\be
e^{ -a^2 {\rm Tr}(U^2+{U^\dagger}^2)}&=&
e^{-2N_f  a^2- a^2 {\rm Tr}(U-{U^\dagger})^2},\nn \\
&=&c e^{-2N_f  a^2}\int d\sigma e^{ {\rm Tr} \sigma^2 /16 a^2+ \frac 12 {\rm Tr}\sigma (U-{U^\dagger})},
\ee
where $\sigma $ is anti-hermitian, and $c$ is a normalization constant.
In a diagonal representation of $\sigma$  denoted by $S \equiv {\rm diag}(is_1,\cdots,
is_{N_f})$
the partition function with
index $\nu$ is thus given by (up to a normalization constant)
\be
Z^\nu_{N_f}(m;a)  &=&e^{-2N_f  a^2} \int d s_k \Delta^2(\{s_k\})  \int_{U(N_f)} \hspace{-1mm} d U \ 
{\det}^\nu U e^{- \sum_k s_k^2/16 a^2}
e^{\frac 12  {\rm Tr}U^\dagger(m+S)+\frac 12  {\rm Tr}U(m-S)  }\nn \\
&=&\int d s_k \Delta^2(\{s_k\})  e^{- \sum_k s_k^2/16 a^2}{\det}^\nu(m- S)\, 
\tilde Z^\nu_{N_f} (\{(m^2 +s_k^2)^{1/2}\}; a=0).
\label{zdiff}
\ee
The Vandermonde determinant is defined by
\be
\Delta(x_1,\cdots, x_p) = \prod_{k>l}^p(x_k-x_l),
\ee
and an explicit expression for the partition function at $a=0$ is given in Eq. (\ref{za=0}).
The expression for the partition function will be discussed in more detail for $N_f =1 $ and 2 in \ref{sec:partition}.

The fermionic partition function has been written as 
an integral over a diffusion kernel times 
the partition function at $a=0$. Next we will show that exactly the same is true for the graded generating functional
obtained in   \cite{DSV}.
The generating function
of the microscopic $p$-point spectral correlation functions
 of the Hermitian Wilson Dirac operator with index $\nu$ is given by
 \be
\label{ZSUSY}
Z^\nu_{N_f+p|p}({\cal M},{\cal Z};a)  & = & \int \hspace{-1.5mm} dU \
{\rm Sdet}(iU)^\nu  
  e^{\frac{i}{2}{\Str}({\cal M}[U-U^{-1}])
    +\frac{i}{2}{\Str}({\cal Z}[U+U^{-1}])
    + a^2{{\Str}(U^2+U^{-2})}},  
\ee
where ${\cal M}\equiv{\rm diag}(m_1\ldots m_{N_f+2p})$ and 
${\cal Z}\equiv{\rm diag}(z_1\ldots z_{N_f+2p})$, and the integration is
over $Gl(N_f+p|p)/U(p)$, see \cite{DOTV}. 
We use the convention that ${\rm
    Trg} \,A={\rm Tr}[A_f]-{\rm Tr}[A_b]$, with $A_f$  the fermion-fermion block of $A$, and $A_b$ 
its boson-boson block. The definition of Sdet follows form the relation ${\rm Sdet} A
= \exp [{\rm Trg\; log A } ]$.
Notice that in comparison to (\ref{Zfull})
the integration over $U$ has been rotated by $i$ so that the convergence
of the bosonic integrals is assured \cite{DSV,ADSVprd}.  
A similar rotation is necessary when computing the spectrum of the 
Dirac operator at nonzero chemical potential \cite{splitchem}. 
The common origin is the non-Hermiticity of the Dirac operator.
The manipulations from (\ref{Zfull}) to (\ref{zdiff}) can be repeated for
the graded partition function. 
We start from the identity
\be
e^{ a^2 {\rm Trg}(U^2+{U^\dagger}^2)}&=&
e^{-2N_f a^2+ a^2 {\rm Trg}(U+U^\dagger)^2},\nn \\
&=&c e^{-2N_f a^2}\int d\sigma e^{ {\rm Trg} \sigma^2 /16 a^2+ \frac i2 {\rm Trg}\sigma (U+{U^\dagger})},
\label{hs}
\ee
 where $\sigma$ is an $(N_f+p|p)$ graded ``Hermitian'' matrix (see (\ref{sigma})) and $c$ is an integration constant.
After shifting integration variables $\sigma \to \sigma -{\cal Z}$ we obtain
(the normalization constants will be fixed at the end of the calculation)
 \be
\label{ZSUSY1}
Z^\nu_{N_f+p|p}({\cal M},{\cal Z};\ha)  & = & e^{-2N_f  a^2}\int d\sigma \int \hspace{-1.5mm} dU \
{\rm Sdet}^\nu (iU ) e^{ {\rm Trg} (\sigma-{\cal Z})^2 /16 a^2
 + \frac i2 {\rm Trg}(\sigma+{\cal M })U
+ \frac i2 {\rm Trg}(\sigma -{\cal M })U^{-1}}
.\nn \\
\ee
We will evaluate this partition function for
 the
$(p+N_f|p)$ graded diagonal matrix 
\be
{\cal Z} ={\rm diag}(\epsilon,\cdots,\epsilon,z_1,\cdots,  z_p, z_1',\cdots, z_p'),
\label{zeta}
\ee
and it  is understood that the limit $\epsilon \to 0$ is taken at the end of the calculation.

The partition function (\ref{Zfull}) and the generating 
function (\ref{ZSUSY}) satisfy the relation
\be
Z_{N_f}^{-\nu}(m,z;a) &=& Z_{N_f}^{\nu}(m,-z;a),\\
Z_{N_f+p|p}^{-\nu}({\cal M},{\cal Z};a) &=& Z_{N_f}^{\nu}({\cal M },-{\cal Z};a).
\ee
For this reason we only consider the case $\nu \ge 0 $ below.   
For $z=0$  (${\cal Z} =0$) the generating function does not depend on the
sign of $\nu$.

The graded matrix $\sigma$ has the structure
\be
\sigma = \mat i\sigma_f & \alpha \\ \beta & \sigma_b \emat
\label{sigma}
\ee
with Hermitian $(p+N_f)\times (p+N_f)$ matrix $ \sigma_f^\dagger = \sigma_f$,  
Hermitian $p\times p$ matrix
$\sigma_b^\dagger = \sigma_b$, and
$\alpha $ and $ \beta$ are matrices  in the Grassmann algebra.
The matrix $\sigma$ can be diagonalized by a super-unitary transformation \cite{Efetov}
\be
\sigma = u S u^{-1}, \quad {\rm with} \quad S\equiv \mat is &0 \\ 0 &
t \emat,
\label{sigma_diag}
\ee
where $s={\rm diag}(s_1,\ldots,s_{p+N_f})$ and $t={\rm
  diag}(t_1,\ldots,t_p)$ contain the real eigenvalues $s_k$ and $t_k$. 

Next we transform to the eigenvalues of $\sigma$ and the super-unitary matrix $U$ as integration variables.
The measure is given by
\be
d\sigma = B_{N_f+p |p}^2(S) \prod_{k=1}^{N_f+p} d s_k \prod_{k=1}^{p} d t_k \,du .
\ee 
with  Berezinian
\be
B_{N_f+p |p}(S) = \frac {\prod_{k>l}^{N_f+p}(is_k-is_l) \prod_{k>l}^p (t_k- t_l)}
{\prod_{k=1}^p\prod_{l=1}^{N_f+p} (t_k-is_l)}.
\ee
The measure $du$ is the superinvariant
Haar measure.

For degenerate quarks masses, ${\cal M}_k = m$, the integral over $u$ can be performed by a graded generalization
\cite{Alfaro-1994,Guhr:1996vx} of the Itzykson-Zuber formula, 
\be
\int du e^{{\rm Trg} (\rho -\xi)^2/2\tau}= B_0
+ \frac{1}{(2\pi\tau)^{(N_f+2p)/2}}  \frac{e^{(1/2\tau){\rm Trg}
    (X^2+R^2)} \det e^{-R_k^b X_l^b/\tau}   
\det e^{-R_k^f X_l^f/\tau} }
{B_{N_f+p|p}(X) B_{N_f+p|p}(R)}.
\ee
Here, $X$ and $R$ are the diagonal representation of $\xi$ and $\rho$, in this order.
$B_0$ contains the contributions due to the boundary terms. They result
from the product of the infinity due  the singularities 
in the measure and
the vanishing result due to the Grassmann integration after changing
to eigenvalues as integration variables. These contributions can be worked
out by expanding the eigenvalues of $\sigma$  in powers of the nilpotent terms (see  \ref{sec:appendix-a}).
They do not contribute to the spectral correlators discussed below and will 
be further analyzed in a future publication.

Ignoring the contribution from $B_0$ we find the generating function
 \be
Z_{N_f+p|p}^\nu({\cal M},{\cal Z}; a)
& =&  \frac{e^{-2N_f  a^2}}{(16\pi a^2)^{p+N_f/2}}
\int \hspace{-1.5mm} dU \
{\rm Sdet}(iU)^\nu 
\int ds dt  \frac{B_{N_f+p|p}(S)}{B_{N_f+p|p}({\cal Z})}
\nn \\
&& 
\hspace*{-3cm}
e^{+ \frac i2 {\rm Trg}(S+m)U
+ \frac i2 {\rm Trg}(S-m)U^{-1}}
\times e^{(1/16 a^2){\rm Trg}(S^2+{\cal Z}^2)} \det e^{-S_k^b
 {\cal Z}_l^b/8 a^2}  \det e^{ -S_k^f {\cal Z}_l^f/8 a^2} .
\label{collect}
\ee
The integral over $U$ can be expressed in terms of the generating function
for $ a = 0$. 
A convenient expression is obtained by observing that
the masses $S_k+m$ and $S_k -m$ can be replaced by the same mass 
$i\sqrt{m^2-S_k^2 }$ for both chiralities. A further simplification
results from the symmetry of the 
integrand  in the $s_k$ and the $t_k$ variables so that all terms in the 
Laplace expansion of the determinants from the Itzykson-Zuber integral give
the same contribution. 
In terms of 
the $a = 0$ partition function we thus obtain
 \be
Z_{N_f+p|p}^\nu(m,{\cal Z}; a)& =& \frac{\prod_{k,l=1}^{p} (z_k'-z_l) \prod_k^{p} {z_k'}^{N_f}}{\Delta(\{z_l \}) \Delta(\{z_k' \}) \prod_k^{p} z_k^{N_f} }
\frac{e^{-2N_f  a^2}}{(16\pi a^2)^{(N_f+2p)/2}} \int ds dt
 {B_{N_f+p|p}(S)\Delta(S/8a^2)}
 \nn \\ &&  \hspace{-2cm} \times   e^{{\rm Trg}[(S-{\cal Z})^2]/16a^2}
 \left ( \frac{\prod_k (m-is_k)}{\prod_{l}(m-t_l)} \right )^{\nu}
 \tilde Z_{N_f+p|p}^\nu\left (\{(s_k^2+m^2)^{1/2}\} ,\{ (m^2-t^2_l)^{1/2}\}
; a= 0 \right ),\nn \\
\label{zfinal}
\ee
where Vandermonde determinant,
$\Delta(S/8a^2)=\Delta(is_1/8a^2,\ldots,is_{N_f}/8a^2)$, and the
prefactors result from the limit $\epsilon \to 0$. 

We have succeeded in rewriting the $a$ dependence of the generating function
as an integral over the product of a diffusion kernel and the $a=0$ generating 
function  given by \cite{SplitVerb1,FA}
 \be
 && \hspace{-2cm}  \tilde Z_{N_f+p|p}^\nu(x_1, \cdots, x_{N_f+2p};a=0) \\
 & = &  c 
\left (\frac{\prod_{k=N_f+p+1}^{N_f+2p} x_k} {\prod_{k=1}^{N_f+p} x_k}\right )^\nu
\frac{\det [ (x_k)^{l-1}{\cal I}_{\nu+l-1}(x_k)]}
{\Delta(x_1^2,\cdots,x_{N_f+p}^2) \Delta(x_{N_f+p+1}^2,\cdots,x_{N_f+2p}^2) }
\nn 
\label{za=0}
\ee
with
\be
{\cal I}_q (x_k) &=& I _q(x_k), \qquad k = 1, \cdots, N_f + p, \nn \\ 
{\cal I}_q (x_k) &=& (-1)^qK _q(x_k), \qquad  k = N_f+p+1, \cdots, N_f +2 p.
\ee
We added a tilde to $  \tilde Z_{N_f+p|p}^\nu$ because the mass-factors due
to the zero modes have been amputated which is not the case
for the partition function at $a\ne 0$.

The $p$ point correlation function is obtained by differentiating with respect
to the $z_k$ and putting $z'_k = z_k$ afterwards. Only if all factors in the
product $\prod(z_k'-z_k) $ are differentiated do we get a nonzero result.
The remaining factors from the Berezinian $B_{N_f+p|p}({\cal Z})$ cancel.
We thus find the correlator of $p$ resolvents
\be
G_{N_f+p|p}^\nu(z_1,\cdots,z_p; a)
& =& \frac {e^{-2N_f  a^2}}{Z_{N_f}^\nu(m;a)}\frac 1{(16\pi a^2)^{(N_f+2p)/2}}
 \int ds dt  {B_{N_f+p|p}(S)\Delta(S/8a^2)} 
\\  && \hspace{-3cm}\times
e^{{\rm Trg}[(S-{\cal Z})^2]/16a^2}\frac{ \prod_{k} (m-is_k)^{\nu}} {\prod_l (m-t_l)^{\nu}}
\tilde Z_{N_f+p|p}^\nu\left (\{ (s_k^2+{m}^2)^{1/2} \},\{ (m^2-t^2_l)^{1/2} \}
; a= 0 \right ).\nn
\label{G_N_f+p|p}
\ee
This result is universal in the sense that it is completely determined
by the symmetries of the QCD partition function.
Below we discuss explicit results for the microscopic spectral density for
$N_f =0, 1$ and $2$. We also check that the partition function for
$N_f$ flavors reduces to previously derived results for $N_f= 0$ and $N_f=1$.

\subsection{Explicit Results for the Spectrum of $D_5$}

In this section we discuss explicit results for the microscopic spectral density of $D_5$ for the quenched case
and one and two dynamical flavors. 

\subsubsection{The quenched case}

In this case the Berezinian is given by
\be
B_{1|1}(S) = \frac 1{t-is}
\ee
and the generating function reduces to
\be
Z^\nu_{1|1}(m,z,z';a) &=&\frac {z'-z}{16a^2\pi}  
\int \frac{ds d t}{t-is} e^{-[(s+iz)^2  +(t-z')^2]/16a^2 }
 \frac{(m-is)^\nu }{(m-t)^\nu } 
\\ && \times 
\tilde  Z^\nu_{1|1}(\sqrt{m^2+s^2},\sqrt{m^2-t^2};a=0).\nn 
\label{genquench}
\ee
Here,
\be
\tilde Z^\nu_{1|1} (x,y;a=0) = \frac {y^\nu}{x^\nu}[y K_{\nu+1}(y)
  I_\nu(x)+xK_\nu(y)I_{\nu +1}(x)]. 
\label{Z11a0}
\ee

The resolvent is given by
\be
G^\nu_{1|1}(z) =  \left . \frac d{dz}\right |_{z'=z} Z^\nu_{1|1}(m,z,z';a).
\ee
Only the term where the prefactor $z'-z$ is differentiated contributes
to the resolvent.
In the microscopic limit, this results in
\be
G^\nu_{1|1}(z) &=& 
-\frac 1{16a^2\pi}\int  \frac{ds d t }{t-is}
e^{-\frac { [s^2  +t^2]}{16a^2} }
 \frac{(m-is-z)^\nu }{(m-t-z)^\nu}\\
&&  \times \tilde Z^\nu_{1|1}(\sqrt{m^2-(is+z)^2},\sqrt{m^2-(t+z)^2};a=0).\nn 
\label{g11}
\ee
Note that we also shifted the integration variables $s$ and $t$ by
$-iz$ and $z'$, respectively. The effect of this shift is discussed in
\ref{sec:appendix-a}. 

The quenched microscopic eigenvalue density of $D_5$ follows from
the imaginary part of the resolvent, cf.~Eq.~(\ref{rho5def}).
 We have checked numerically that this result coincides with the
 result obtained from a standard supersymmetric computation in
 \cite{DSV}. See \cite{DSV,ADSVprd} for plots of the quenched density.

In the $a\to 0$ limit at fixed $m$ and $z$, the Gaussian integrals in Eq. (\ref{genquench}) 
become $\delta$-functions
which can be integrated resulting in
\be
\lim_{a\to 0} Z^\nu_{1|1}(m,z,z';a) = \left [\frac{(z-m)}{(z'-m)  }\right ]^\nu
\tilde Z^\nu_{1|1} (\sqrt{m^2-z^2}, \sqrt{m^2-{z'}^2};a=0).
\ee 
The prefactor gives
the contribution from the zero modes to the resolvent
\be
\frac \nu{z-m},
 \ee
whereas the second factor gives the contribution of the nonzero modes
for $a= 0$. 


\subsection{One Flavor}

For $N_f=1$ the generating function is given by
\be
Z^\nu_{2|1} &=&  \frac{e^{-2a^2}}{64 a^3 \pi^{3/2}} \frac{(z'-z)z'}z 
\int \frac {dt ds_1 ds_2 (is_2-is_1) }{(t-is_1)(t-is_2)}
e^{-[ s_1^2 +(s_2+iz)^2  +(t-z')^2]/16a^2   }
 \\
&&\times
\frac{(m-is_1)^\nu(m-is_2)^\nu}{(m-t)^\nu}
\tilde Z^\nu_{2|1}( (s_1^2+m^2)^{1/2},
{(s_2^2+m^2)^{1/2}},(m^2-t^2)^{1/2};a=0),\nn
\label{z21}
\ee
where
\be
\tilde Z_{2|1}^\nu(x_1,x_2,x_3 ) =  \frac {x_3^\nu}{x_1^\nu x_2^\nu(x_2^2 -x_1^2)}
\det \left ( 
\begin{array}{ccc}
I_\nu(x_1)& x_1I_{\nu+1}(x_1) & x_1^2 I_{\nu+2}(x_1)   \\ 
I_\nu(x_2)& x_2I_{\nu+1}(x_2) & x_2^2 I_{\nu+2}(x_2) \\ 
(-1)^\nu K_\nu(x_3)& (-1)^{\nu+1} x_3K_{\nu+1}(x_3) & (-1)^{\nu} x_3^2 K_{\nu+2}(x_3)    
\end{array} \right ).\nn \\
\ee

To obtain the resolvent we differentiate with respect to $z'$ and put 
$z'=z$ after differentiation. An additional minus sign arises because $z'$
is the bosonic source term. We thus find 
\be
G^\nu_{2|1}(z) &=& - \frac{e^{-2a^2}}{64a^3 \pi^{3/2} Z^\nu_1(m;a)} 
\int  \frac {ds_1ds_2dt(is_2-is_1) }{(t-is_1)(t-is_2)}
e^{-[ s_1^2 +(s_2+iz)^2  +(t-z)^2]/16a^2   }
 \\
&&\times
\frac{(m-is_1)^\nu(m-is_2)^\nu}{(m-t)^\nu}
\tilde Z^\nu_{2|1}( (s_1^2+m^2)^{1/2},
{(s_2^2+m^2)^{1/2}},(m^2-t^2)^{1/2};a=0),\nn
\ee
where the $s_2$ and $t$ integration contours are shifted such that $s_2+iz$ and $t-z$
run over the real axis.
The resolvent is normalized with respect to the one flavor partition function
given in Eq. (\ref{zone}). As for the quenched case we have checked 
numerically that the one flavor microscopic density, which follows
from the imaginary part of $G^\nu_{2|1}$, is identical to  the
result \cite{ADSVNf1} obtained from chiral perturbation theory
using the standard supersymmetric method. We refer to \cite{ADSVNf1} for plots.

The small $a$ limit for fixed $m$ and $z$  can be obtained by first shifting $s_2 \to s_2-iz$
and $t \to t +z $ and then expand the nonexpential factors in 
Eq. (\ref{z21}). The integration measure can be expanded as
\be
 \frac {(is_2-is_1+z) }{(t+z-is_1)(t-is_2)} =[ 1 +\frac 1z (is_2-t)]\frac 1{t-is_2} + \cdots.
\label{berexp}
\ee
For small $a$ the partition function $ \tilde Z_{2|1}((s_1^2+m^2)^{1/2},
{((s_2-iz)^2+m^2)^{1/2}},(m^2-(t+z)^2)^{1/2};a=0)$ (denoted by
$\tilde Z_{2|1}$ below) 
can be expanded 
to first order in $s_k$ and $t$
\be
&& m^{-\nu} \tilde Z_1^\nu(m; a=0) 
+ s_1 \left .\frac d{ds_1}\tilde Z_{2|1}\right |_{s_1=s_2=t =0} 
+ s_2 \left .\frac d{ds_2}\tilde Z_{2|1}\right |_{s_1=s_2=t =0} 
+ t \left .\frac d{dt}\tilde Z_{2|1}
\right |_{s_1=s_2=t =0}.\hspace*{1cm}
\ee
The term linear in $s_1$ does not contribute to leading order in $a$ and 
the remaining terms can be written as
\be
&& m^{-\nu} \tilde Z_1^\nu(m; a=0)  + (t-is_2) \left .\frac d{dt}\tilde Z_{2|1}
\right |_{s_1=s_2=t =0}\nn\\
&&= m^{-\nu} \tilde Z_1^\nu(m; a=0)  + (t-is_2) \left .\frac d{dz'}\right |_{z'=z}
\tilde Z_{2|1}(m,\sqrt{m^2-z^2},\sqrt{m^2-{z'}^2}).
\label{nonzeromodes}
\ee
The linear term in $s_1$ in
 the expansion of the prefactor
\be
\frac{(m-is_1)^\nu(m-z-is_2)^\nu}{(m-z-t)^\nu}
= \frac {m^\nu(m-z)^\nu}{(m-z')^\nu}
\left [  -\nu\frac{is_1}m  +\nu\frac{t-is_2}{m-z} \right ]+\cdots.
\label{zeromodes}
\ee
also vanishes after integration.
Combining the contributions from Eqs. (\ref{berexp}-\ref{zeromodes})
we obtain
the small $a$ limit 
\be
G^\nu(z) = \frac \nu{z-m} +\tilde G^\nu(m,z;a=0) + \frac 1z,
\ee
where
$\tilde G^\nu(m,z;a=0)$ is the resolvent of the nonzero eigenvalues.
The additional $1/z$ term will be canceled by 
Efetov-Wegner terms which contribute to the real part of the resolvent.


\subsection{Two Flavors}

In this section we write out the $N_f =2 $ generating function with dynamical
quark masses $m$ and index $\nu$. With the Berezinian given by
\be
B_{3|1}(s) = \frac{(is_2-is_1)(is_3-is_1)(is_3-is_2)}{(t-is_1)(t-is_2)(t-is_3)}
\ee
we find the generating function
\be
\label{Z31}
Z^\nu_{3|1} &=& -\frac {e^{-4a^2}}{\pi^2(16a^2)^3}\frac{z'^2(z'-z)}{z^2}
\int\frac{ ds_1 ds_2 ds_3 dt (is_2-is_1)^2(is_3-is_1)(is_3-is_2)} 
{(t-is_1)(t-is_2)(t-is_3)}\nn\\ && \times
e^{-[ s_1^2+s_2^2 +(s_3+iz)^2 +(t-z')^2] /16a^2 }
\frac{(m-is_1)^\nu(m-is_2)^\nu(m-is_3)^\nu}{(m-t)^\nu}
\nn\\ && \times
\tilde Z^\nu_{3|1}((s_1^2+m^2)^{1/2},(s_2^2+m^2)^{1/2},(s_3^2+m^2)^{1/2},
(m^2-t^2)^{1/2}; a=0).
\ee
The partition function for $a=0$ is given by
\be
\label{Z31a0}
\tilde Z^\nu_{3|1}(x_1,x_2,x_3,x_4;a=0) &=& 2 \frac{x_4^\nu}{x_1^\nu x_2^\nu x_3^\nu} 
\frac 1{(x_3^2-x_2^2)(x_3^2-x_1^2)(x_2^2-x_1^2)}
\\&& \hspace*{-6cm}\times
\det \left ( \begin{array}{cccc}
I_\nu(x_1)& x_1I_{\nu+1}(x_1) & x_1^2 I_{\nu+2}(x_1) & x_1^3 I_{\nu+3}(x_1) \\   
I_\nu(x_2)& x_2I_{\nu+1}(x_2) & x_2^2 I_{\nu+2}(x_2) & x_2^3 I_{\nu+3}(x_2)  \\ 
I_\nu(x_3)& x_3I_{\nu+1}(x_3) & x_3^2 I_{\nu+2}(x_3) & x_3^3 I_{\nu+3}(x_3)  \\ 
(-1)^\nu K_\nu(x_4)& x_4(-1)^{\nu+1}K_{\nu+1}(x_4) & x_4^2 (-1)^{\nu+2} K_{\nu+2}(x_4) & x_4^3 (-1)^{\nu+3} K_{\nu+3}(x_4)   
\end{array} \right ).\nn
\ee

\begin{figure}
\centerline{\includegraphics[width=8cm,clip]{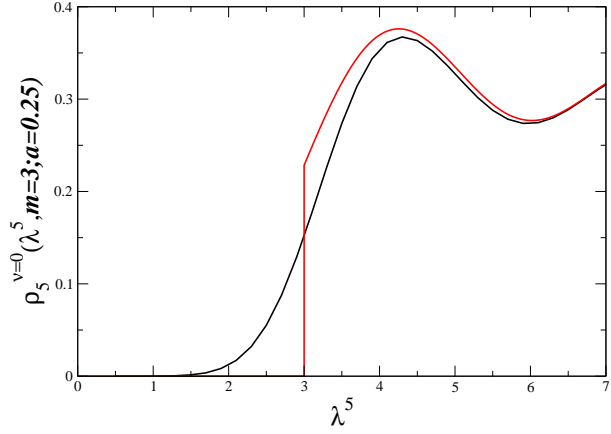}}
\noindent
\caption{Spectral density of $D_5$ for $N_f=2$, $\nu=0$, 
$m= 3$ for  $a=0.25$ (black curve) and $a = 0$ (red curve).
\label{fig:two}} 
\end{figure}

\begin{center}
\begin{figure}
\includegraphics[width=8.5cm,angle=0]{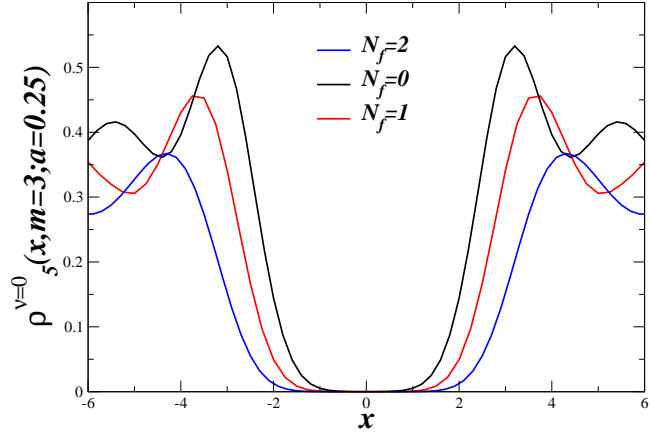}
\caption{\label{fig:rhoNf012} The spectral density of the
  eigenvalues of $D_5$ in the sector with index $\nu=0$  
  plotted for $N_f=0,1$ and $2$. The value of $a=0.25$ and $m=3$. 
  The increasing repulsion from the origin for larger $N_f$ is clearly
  visible.} 
\end{figure}
\end{center}

The resolvent is obtained by differentiating the factor $z'-z$ with
respect to $z$ at $z'=z$. This results in
\be
G^\nu_{3|1}(z;a) &=& \frac {e^{-4a^2}}{\pi^2(16a^2)^3Z^\nu_{N_f =2}(m;a)}
\int ds_1 ds_2 ds_3 dt \frac{(is_2-is_1)^2(is_3-is_1)(is_3-is_2)} 
{(t-is_1)(t-is_2)(t-is_3)}\nn \\ &&\times
e^{-\frac 1{16a^2}[ s_1^2+s_2^2 +(s_3+iz)^2 +(t-z)^2]  }
\frac{(m-is_1)^\nu(m-is_2)^\nu(m-is_3)^\nu}{(m-t)^\nu}
\nn\\&&\times
\tilde Z^\nu_{3|1}((s_1^2+m^2)^{1/2},(s_2^2+m^2)^{1/2},(s_3^2+m^2)^{1/2},
(m^2-t^2 )^{1/2};a=0),
\label{G31}
\ee
where the $s_3$ and $t$ integration contours are shifted such that $s_3+iz$ and $t-z$
run over the real axis.
The resolvent has been normalized with respect to 
the two-flavor partition function  defined in Eq. (\ref{ztwo}).

The microscopic eigenvalue density of $D_5$ with two light flavors of
mass $m$ is then given by
\be
\rho^{\nu,N_f=2}_5(\lambda^5,m;a) = \frac{1}{\pi}{\rm Im}[G^\nu_{3|1}(-\lambda^5)]_{\epsilon\to0}.
\label{rho5_Nf2}
\ee
In Fig. \ref{fig:two} we show the two-flavor microscopic spectral density
of $D_5$ as a function of $\lambda^5$ for $\nu = 0$ and $m=3$
and compare the result for $ a=0.25$ and $a=0$. 
The area below the two curves is the same within our numerical accuracy.
In Fig. \ref{fig:rhoNf012} we compare the two-flavor result to the
one-flavor result and the quenched result with the same parameters.

\begin{center}
\begin{figure}[t!]
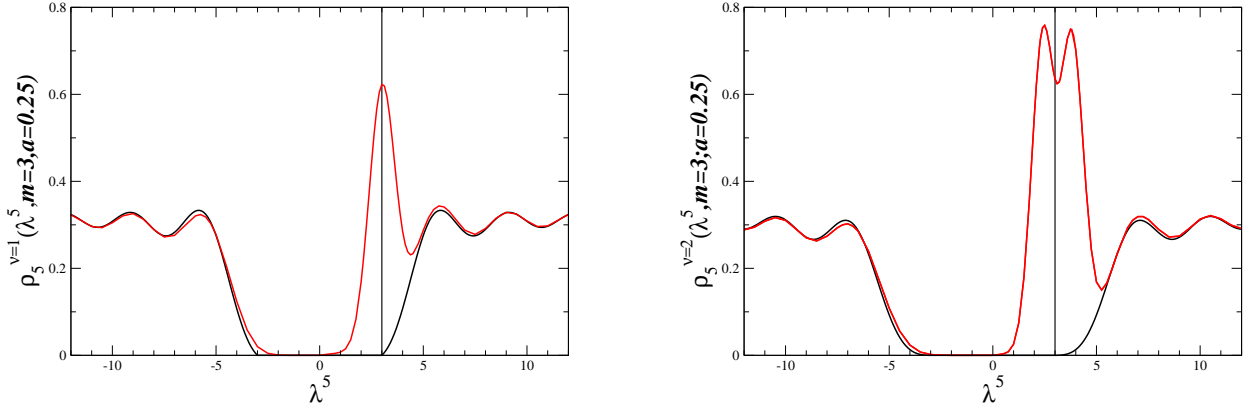

\includegraphics[width=7.5cm,angle=0]{rho5_Nf2_a0.25PRL_m3_nu1.eps}
\hfill
\includegraphics[width=7.5cm,angle=0]{rho5_Nf2_a0.25_m3_nu2.eps}
\caption{\label{fig:Nf2nu} The effect of the index on the spectral
  density of the Hermitian Wilson Dirac operator for two
  flavors. Results for index $\nu=1$ (left) and  $\nu=2$ (right) 
are shown for $m=3 $ and 
  $a=0.25$ (red curve) and $a=0$ (black curve). 
The vertical black
  line marks the position of the $\nu$ fold $\delta$-function due to
  the exact topological modes at $a=0$. Note that the primary effect
  of $a$ for small $a$ is to smear out the $\delta$-function.}
\label{fig:nf2nu}
\end{figure}
\end{center}

The effect of non-zero index $\nu$ for $N_f=2$ is displayed in
Fig.~\ref{fig:Nf2nu}. Note that the two flavor eigenvalue density is 
positive definite (the square of the Wilson-fermion determinant is real
and positive) and that the spectral gap cannot close completely
on the microscopic scale due to the repulsion from the origin 
(the square of the Wilson-fermion determinant vanishes quadratically
as an eigenvalue of $D_5$ approaches zero).   
It would be most interesting to compare these 
analytical predictions to dynamical lattice data, 
such as those presented in \cite{Luscher}.

For small $a$ at fixed $m$ and $z$ we can write
\be
\frac{(is_3-is_1+z)(is_3-is_2+z)} 
{(t-is_1+z)(t-is_2+z)(t-is_3)} = [1 + \frac 2z (is_3-t) ] \frac 1{t-is_3}.
\ee
The constant term contributes to the real part of the
resolvent, and as in the one-flavor case, we expect that it will be canceled
by contributions from the Efetov-Wegner terms. The $a \to 0$ limit of the
$N_f =2$ resolvent is obtained by expanding the pre-exponential factors
in Eq. (\ref{G31}) as in the one-flavor case.


\section{ The distribution of real modes}

We now consider the eigenvalues of the usual Wilson Dirac operator $D_W$. 
For small nonzero values of the lattice 
spacing $a$, the eigenvalues, $\lambda^W$, of $D_W$ spread into a narrow 
band around the imaginary axis of the complex eigenvalue plane. The 
eigenvalues in the complex plane make up complex conjugate pairs
$\lambda^W,(\lambda^W)^*$ or are exactly real. 
See \cite{Itoh} for a derivation of these properties of the Wilson Dirac
operator. In this section we analyze the microscopic spectral
correlation functions for the real eigenvalues of $D_W$.

The generating function for the $p$-point correlation function with
$N_f$ dynamical quarks in the sector of gauge field configurations
with index $\nu$ takes the form
\be
Z^\nu_{N_f+p|p} = \left\langle {\det}^{N_f} (D_W + m_f)
\prod_{k=1}^p\frac {\det (D_W +m_k)}
{\det (D_W +m_k'-i\epsilon\gamma_5)}\right\rangle.
\label{zrealdef}
\ee
The spectral resolvent for the one point function is 
\be
\label{S1susy}
\Sigma^\nu_{N_f+1|1}(m,m_f;a)= \lim_{m'\to m} \frac{d}{dm} {\cal Z}^\nu_{N_f+1|1}(m_f,m,m';a).
\ee
To be precise, the one point function that corresponds to this resolvent
is the distribution of the
chiralities, ${\rm sign}(\langle k| \gamma_5 | k \rangle)$, over the
real modes, $\lambda_k^W\in {\mathbb R}$, 
\be
\label{distOFchi}
\rho^\nu_\chi(\lambda^W,m_f;a) 
& \equiv & \left\langle\sum_{\lambda_k^W\in {\mathbb
    R}}\delta(\lambda_k^W +\lambda^W) \, {\rm sign}(\langle k|
\gamma_5 | k \rangle)\right\rangle_{N_f}. 
\ee
It can be obtained from the discontinuity of the spectral resolvent across the
real axis (see section II of \cite{ADSVprd})
\be
\label{discSigma}
\rho^\nu_\chi(\lambda^W,m_f;a) 
 & =& \frac{1}{\pi} {\rm Im}[\Sigma^\nu_{N_f+1|1}(m_f,m=\lambda^W-i\epsilon;a)]_{\epsilon\to0}.
\ee
The $p$-point spectral resolvent is given by
\be
\label{Spsusy}
\Sigma^\nu_{N_f+p|p}(m_1,\ldots,m_p,m_f;a) &= &\lim_{m'_1\to
  m_1}\ldots\lim_{m'_p\to m_p} \frac{d}{dm_1}\cdots\frac{d}{dm_p} \\
 && \hspace{1cm} \times {\cal Z}^\nu_{N_f+p|p}(m_f,m_1,\ldots,m_p,m'_1,\ldots,m'_p;a).\nn
\ee
As in the case of the one-point function, the discontinuities across the real axis
give the $p$-point density correlation functions.

As was discussed in \cite{DSV}, the generating function for the correlation functions
(\ref{Spsusy}) is given by
\be
\label{Zreal}
Z^\nu_{N_f+p|p}({\cal M};a)  & = & \int \hspace{-1.5mm} dU \
{\rm Sdet}(iU)^\nu  
  e^{\frac{i}{2}{\Str}({\cal M}[U-U^{-1}])
    + a^2{{\Str}(U^2+U^{-2})}}, 
\ee
which is just the generating function (\ref{ZSUSY}) for $ {\cal Z} = 0$.
The mass matrix corresponding to (\ref{zrealdef}) is given by
the
$(p+N_f|p)$ graded diagonal matrix 
\be
{\cal M}\equiv {\rm diag}(m_f,\cdots,m_f,m_1,\cdots, m_p,m_1',\cdots,m_p').
\ee
The first $N_f$ entries are the physical masses and need not be identical.

In order to derive the $p$-point function the we start with
 the identity (instead of the identity (\ref{hs}))
\be
e^{ a^2 {\rm Trg}(U^2+{U^{-1}}^2)}&=&
e^{2N_f a^2+ a^2 {\rm Trg}(U-U^{-1})^2},\nn \\
&=&c e^{2N_f  a^2}\int d\sigma e^{ {\rm Trg} \sigma^2 /16 a^2+ \frac i2 {\rm Trg}\sigma (U-{U^{-1}})},
\label{hsreal}
\ee
where $\sigma$ is an $(N_f+p|p)$ graded Hermitian matrix (see Eq. (\ref{sigma})) and $c$ a normalization constant..
After a  shift of $\sigma$ by ${\cal M}$ we obtain 
 \be
\label{Zreal1}
Z^\nu_{N_f+p|p}({\cal M};a)  & = & e^{2N_f  a^2}\int d\sigma \int \hspace{-1.5mm} dU \
{\rm Sdet}^\nu(i U)  e^{ {\rm Trg} (\sigma-{\cal M})^2 /16 a^2
 + \frac i2 {\rm Trg}\,\sigma(U -U^{-1})}.
\ee
The next step is to decompose $\sigma = u S u^{-1}$ with $S$ a diagonal graded matrix
(see Eq. (\ref{sigma_diag}))
and perform the integration
over $u$ by a supersymmetric generalization of the Itzykson-Zuber integral. 
We find
\be
Z^\nu_{N_f+p|p}({\cal M};a)
& =&  \frac{e^{2N_f  a^2}}{(16\pi a^2)^{(N_f+2p)/2}}\int ds dt \frac {B_{N_f+p|p}(S)}{B_{N_f+p|p}({\cal M})} 
\det{e^{-{\cal M}^b_k S^b_l/8a^2}}\det{e^{-{\cal M}^f_k S^f_l/8a^2}}
 \nn\\  && 
  \times  e^{{\rm Trg}[S^2 +{\cal M}^2]/16a^2}\frac{ \prod_{k} (is_k)^{\nu}} {\prod_l (t_l)^{\nu}}
 \tilde Z_{N_f+p|p}^\nu\left ( \Big\{\sqrt{(is_k)^2}
 \Big\},\Big\{\sqrt{(t_l)^2} \Big\}; a=0 \right).\hspace*{0.5cm}
\label{S_N_f+p|p}
\ee
For degenerate dynamical quarks the above expression can be further simplified.
\be
\det e^{-{\cal M}^f_k S^f_l/8a^2} &=& \Delta(m_1,\cdots, m_{N_f})
  \Delta(S_1^f/8a^2,\cdots, S_{N_f}^f/8a^2)e^{-m(S_1^f+\cdots +S_{N_f}^f)/8a^2}
\nn \\ && \times{\det[ e^{-m_kS_l^f/8a^2}]}_{k,l = N_f+1, \cdots, N_f+p}
 + \quad {\rm permutations\; of}\;\; S_k^f.
\ee
All permutations of the $S_k$ give the same contribution. For $m'_k \to m_{N_f+k}$ we
obtain
\be
\frac{\Delta(m_1,\cdots, m_{N_f})}{B_{N_f+p|p}({\cal M})} \to
\prod_{k=1}^p(m'_k -m_{N_f+k}).
\ee
The final expression for the generating function with degenerate quark masses
is given by
\be
Z^\nu_{N_f+p|p}({\cal M};a)
& =&  
\frac{\prod_{k=1}^p(m'_k -m_{N_f+k})e^{2N_f  a^2}}{(16\pi a^2)^{(N_f+2p)/2}}\int ds dt {B_{N_f+p|p}(S)}
 \Delta(S_1^f/8a^2,\cdots, S_{N_f}^f/8a^2)
\nn\\  && \hspace*{-1cm}
  \times  e^{(1/16a^2){\rm Trg}[(S-{\cal M})^2]}\frac{ \prod_{k} (is_k)^{\nu}} {\prod_l (t_l)^{\nu}}
\tilde  Z_{N_f+p|p}^\nu\left ( \Big\{\sqrt{(is_k)^2}
 \Big\},\Big\{\sqrt{(t_l)^2} \Big\}; a=0 \right).\hspace*{1cm}
\label{S_N_f+p|p-2}
\ee
In order to obtain nonzero contributions to the spectral resolvent
(\ref{Spsusy}) all $m_k$ in the pre-factor have to be differentiated.
Below we give the explicit expressions in a couple of cases relevant
for current lattice simulations.

\subsection{The quenched case}

The quenched one-point function $\rho^\nu_\chi(\lambda^W;a)$ follows from 
\be
\Sigma^\nu_{1|1}(m;a) &=& 
-\frac 1{16a^2\pi}\int  \frac{ds dt}{t-is}
e^{-\frac {s^2+t^2}{16a^2}}
 \frac{(is+m)^\nu}{(t+m)^\nu}\\
&& \hspace{3cm} \times  \tilde Z^\nu_{1|1}(\sqrt{(is+m)^2},\sqrt{(t+m)^2};a=0),\nn 
\label{S11}
\ee
after using (\ref{discSigma}). The explicit form of $Z^\nu_{1|1}$ at $a=0$
is given in Eq.~(\ref{Z11a0}).

\vspace{2mm}

The two-point function in the quenched case follows from the
discontinuity of  
\be\label{S22shift}
\Sigma^\nu_{2|2}(m_1,m_2;a) &=& 
\frac 1{(16 \pi a^2)^2} 
\int ds_1 ds_2 dt_1 dt_2 \nn\\
&&\hspace{-1cm}\times\frac{(is_2+m_2-is_1-m_1)(t_2+m_2-t_1-m_1)} 
{(t_1-is_1)(t_2+m_2-is_1-m_1)(t_1+m_1-is_2-m_2)(t_2-is_2)}\nn \\ &&\hspace{-1cm}\times
e^{-\frac 1{16a^2}[s_1^2+s_2^2+t_1^2+t_2^2]}
\frac{(is_1+m_1)^\nu(is_2+m_2)^\nu}{(t_1+m_1)^\nu(t_2+m_2)^\nu}
\\&&\hspace{-1cm}\times
\tilde Z^\nu_{2|2}(\sqrt{(is_1+m_1)^2},\sqrt{(is_2+m_2)^2},\sqrt{(t_1+m_1)^2},
\sqrt{(t_2+m_2)^2};a=0)\nn
\ee
across the real $m_1$ and $m_2$ axis. Here the $a=0$ partition
function takes the form
\be
\label{Z22a0}
\tilde Z^\nu_{2|2}(x_1,x_2,x_3,x_4;a=0) &=& 2\frac{x_3^\nu x_4^\nu}{x_1^\nu x_2^\nu } 
\frac 1{(x_2^2-x_1^2)(x_4^2-x_3^2)}
\\&& \hspace*{-6cm}\times
\det \left ( \begin{array}{cccc}
I_\nu(x_1)& x_1I_{\nu+1}(x_1) & x_1^2 I_{\nu+2}(x_1) & x_1^3 I_{\nu+3}(x_1) \\   
I_\nu(x_2)& x_2I_{\nu+1}(x_2) & x_2^2 I_{\nu+2}(x_2) & x_2^3 I_{\nu+3}(x_2)  \\ (-1)^\nu K_\nu(x_3)& x_3(-1)^{\nu+1}K_{\nu+1}(x_3) & x_3^2 (-1)^{\nu+2} K_{\nu+2}(x_3) & x_3^3 (-1)^{\nu+3} K_{\nu+3}(x_3)
  \\ 
(-1)^\nu K_\nu(x_4)& x_4(-1)^{\nu+1}K_{\nu+1}(x_4) & x_4^2 (-1)^{\nu+2} K_{\nu+2}(x_4) & x_4^3 (-1)^{\nu+3} K_{\nu+3}(x_4)   
\end{array} \right ).\nn
\ee 

The two-point correlation function contains a term due to self-correlations,
\be
R_2(x,y)  = \left \langle \sum_{k,l}\frac 1{x+\lambda_k} \frac1{y+\lambda_l} \right\rangle
=  \left \langle \sum_k  \frac 1{x+\lambda_k} \frac 1{y+\lambda_k}\right \rangle 
+\left \langle  \sum_{k \ne l} \frac 1{x+\lambda_k} \frac 1{y+\lambda_l} \right \rangle .
\ee
This term  can be rewritten as
\be
\left \langle \sum_k \frac  1{x+\lambda_k} \frac 1{y+\lambda_k} \right \rangle=
\frac 1{y-x}  \left \langle \sum_k  \frac 1{x+\lambda_k}   
- \frac1{y+\lambda_k}  \right\rangle 
\ee
which is singular for $y \to x$ if $x$ and $y$ are on opposite sides of the cut of the resolvent.
It can be shown in general terms \cite{mario} that such singular terms are due to Efetov-Wegner terms
and are not included in the expression (\ref{S22shift}). They will be analyzed in a future publication.
The two-point spectral correlation function  
\be
\rho^\nu_\chi(\lambda^W_1,\lambda^W_2;a)&=&\left\langle\sum_{\lambda_k^W,\lambda_l^W\in {\mathbb
    R}}\delta(\lambda_k^W +\lambda^W_1) \, {\rm sign}(\langle k|
\gamma_5 | k \rangle)\delta(\lambda_l^W +\lambda^W_2) \, {\rm sign}(\langle l|
\gamma_5 | l \rangle)\right\rangle \\
&&-\left\langle\sum_{\lambda_k^W\in {\mathbb
    R}}\delta(\lambda_k^W +\lambda^W_1) \, {\rm sign}(\langle k|
\gamma_5 | k \rangle)\right\rangle\left\langle\sum_{\lambda_k^W\in {\mathbb
    R}}\delta(\lambda_k^W +\lambda^W_2) \, {\rm sign}(\langle k|
\gamma_5 | k \rangle)\right\rangle\nn
\ee
can also be decomposed into  
sum of self-correlations and genuine two-point correlations
\be
\rho^\nu_\chi(\lambda^W_1,\lambda^W_2;a)&=&
\delta(\lambda^W_1 -\lambda^W_2)\left\langle\sum_{\lambda_k^W\in {\mathbb
    R}}\delta(\lambda_k^W +\lambda^W_1) \right\rangle \\
&& +\left\langle\sum_{\lambda_k^W,\lambda_l^W\in {\mathbb
    R},k\neq l}\delta(\lambda_k^W +\lambda^W_1) \, {\rm sign}(\langle k|
\gamma_5 | k \rangle)\delta(\lambda_l^W +\lambda^W_2) \, {\rm sign}(\langle l|
\gamma_5 | l \rangle)\right\rangle \nn\\
&&-\left\langle\sum_{\lambda_k^W\in {\mathbb
    R}}\delta(\lambda_k^W +\lambda^W_1) \, {\rm sign}(\langle k|
\gamma_5 | k \rangle)\right\rangle\left\langle\sum_{\lambda_k^W\in {\mathbb
    R}}\delta(\lambda_k^W +\lambda^W_2) \, {\rm sign}(\langle k|
\gamma_5 | k \rangle)\right\rangle.\nn
\ee
An important observation is that the sign of the chirality drops out in the expression
for the self-correlations so
that the diagonal part of the two-point correlator gives the density of real modes
\be
\rho^\nu_\chi(\lambda^W_1,\lambda^W_2=\lambda^W_1;a)&=&
\left\langle\sum_{\lambda_k^W\in {\mathbb
    R}}\delta(\lambda_k^W +\lambda^W_1) \right\rangle.
\ee
 
\subsection{One dynamical flavor}

With one dynamical flavor of mass $m_f$ we have 
\be
\Sigma^\nu_{2|1}(m,m_f;a) &=& - \frac{e^{2a^2}}{64a^3 \pi^{3/2} Z^\nu_1(m_f;a)} 
\int  \frac {ds_1ds_2dt(is_2+m-is_1-m_f) }{(t+m-is_1-m_f)(t-is_2)}
e^{-[s_1^2+s_2^2+t^2]/16a^2   }
 \nn \\
&&\hspace*{-2cm}\times
\frac{(-is_1-m_f)^\nu(is_2+m)^\nu}{(t+m)^\nu}
\tilde Z^\nu_{2|1}(\sqrt{(is_1+m_f)^2},
\sqrt{(is_2+m)^2},\sqrt{(t+m)^2};a=0).\nn\\
\ee
Note that the one flavor theory has a sign problem and consequently
the one-point function 
\be
\rho^\nu_\chi(\lambda^W,m_f;a)=\frac{1}{\pi}{\rm Im}[\Sigma^\nu_{2|1}(m-i\epsilon,m_f;a)]_{\epsilon\to0}
\ee
changes sign at $\lambda^W=m_f$, see also \cite{ADSVNf1} where this
function was derived by a direct supersymmetry computation.  

The $a\to 0$ limit atr fixed $m$ and$m_f$ is obtained by expanding the pre-exponential factors
to first order in the $s_k$ and $t$. This results in
\be
\Sigma_{2|1}^\nu(m, m_f;a) =\frac \nu m + \frac 1{m-m_f} +\Sigma_{2|1}(m,m_f;a=0).
+ \cdots .
\ee
The $1/(m-m_f)$ term is expected to cancel against the  Efetov-Wegner terms.


\subsection{Two dynamical flavors}

Finally we give the explicit form of the distribution of the
chiralities over the real eigenvalues of $D_W$ in a sector with fixed
index $\nu$ for the physically relevant case of two light flavors with
mass $m_f$. The spectral resolvent can be expressed as  
\be
\Sigma^\nu_{3|1}(m,m_f;a) &=& \frac {e^{4a^2}}{\pi^2(16a^2)^3Z^\nu_{N_f =2}(m_f;a)}
\int ds_1 ds_2 ds_3 dt 
\nn\\
&&\hspace{-1cm}\times\frac{(is_2-is_1)^2(is_3+m-is_1-m_f)(is_3+m-is_2-m_f)} 
{(t+m-is_1-m_f)(t+m-is_2-m_f)(t-is_3)}\nn \\ &&\hspace{-1cm}\times
e^{-\frac 1{16a^2}[ s_1^2+s_2^2 +s_3^2 +t^2]  }
\frac{(is_1+m_f)^\nu(is_2+m_f)^\nu(is_3+m)^\nu}{(t+m)^\nu}
\\&&\hspace{-1cm}\times
\tilde Z^\nu_{3|1}(\sqrt{(is_1+m_f)^2},\sqrt{(is_2+m_f)^2},\sqrt{(is_3+m)^2},
\sqrt{(t+m)^2};a=0),\nn
\label{S31shift}
\ee
where the two flavor partition function in the prefactor is given
by Eq. (\ref{ztwo}) and the explicit form of $Z^\nu_{3|1}$ at $a=0$ is given
in (\ref{Z31a0}).


\section{Conclusions}
\label{sec:conclusions}

We have obtained analytical expressions for all microscopic 
spectral correlation functions of the Wilson Dirac operator for any
number of flavors with equal quark mass.
In particular, we have
computed the microscopic spectrum of the Hermitian Wilson
Dirac operator in the physically relevant two flavor case
and the distribution of the chiralities over the real eigenvalues
of the Wilson Dirac operator. The results 
were obtained from a chiral Lagrangian for the generating function of
the Wilson-Dirac spectrum using the graded eigenvalue method.
We have also given expressions for an arbitrary number of flavors as
well as higher order correlation functions. 
We have checked that our results for zero and one flavor are in complete agreement
with a previous calculation based on a brute force
supersymmetric computation. 
Since these results are based on a chiral Lagrangian that follows from the
global symmetries of the lattice QCD partition function they can
also be derived from a chiral random matrix theory for 
the Wilson Dirac operator with the same global symmetries. This enables
us to derive additional results using random matrix techniques which we
hope to address in a future paper.

 The new results give the leading order effect of the lattice
discretization on the spectrum of the Wilson Dirac operator also
in the physically relevant two flavor case.
The analytical understanding of the smallest eigenvalues of
the Wilson Dirac operator  can be used to optimize the choices of
parameters in lattice QCD for which the simulation is stable. Our
results also offer a direct way to measure the low energy constants
of Wilson chiral perturbation theory.

\noindent
{\bf Acknowledgments:}
We would like to thank Thomas Guhr for encouraging us to apply the graded
eigenvalue method to this problem. We also acknowledge Gernot Akemann,
Poul Henrik Damgaard and  Mario Kieburg for useful discussions and for a
critical reading of the manuscript.
This work was supported  by U.S. DOE Grant No. 
DE-FG-88ER40388 (JV) and the {\sl Sapere Aude} program of The Danish
Council for Independent Research (KS).  

\renewcommand{\thesection}{Appendix \Alph{section}}
\setcounter{section}{0}

\section{ Efetov-Wegner Terms}

\label{sec:appendix-a}
In this Appendix we illustrate the effect of Efetov-Wegner terms for
the Gaussian super-integral
\be
Z(z_1,z_2) =\int d\sigma e^{{\rm Trg} (\sigma -{\cal Z})^2/16a^2},
\label{zgdef}
\ee
where $\sigma$ and ${\cal Z}$ are  the $(1|1) $ supermatrices
\be
 \sigma = \mat a & \chi \\ \rho & b \emat,\qquad
{\cal Z} = \mat z_1 & 0 \\ 0& z_2 \emat,
\ee
and $d\sigma$ is the integral over the matrix elements of $\sigma$. Clearly,
the integral does not depend on $z_1$ and $z_2$ so that after a proper
normalization of the measure we have
\be
Z(z_1,z_2) = 1.
\label{znorm}
\ee
The supermatrix $\sigma$ can be diagonalized by
\be
\sigma = u \mat is & 0 \\ 0 & t \emat  u^{-1}
\label{diag}
\ee
with
\be
u = \exp\mat 0 & \alpha \\ \beta  & 0 \emat.
\ee 
We first perform the integral by transforming to the eigenvalues of
$\sigma$ as integration variables and then perform the integral over $u$ by
a supersymmetric generalization of the Itzykson-Zuber integral. This
results in (Actually this is a special case of Eq. (\ref{collect}) where
the partition function for $a = 0$ is put equal to unity.)
\be
Z(z_1,z_2) = \frac{(z_2-z_1)}{16\pi a^2} \int_{-\infty}^\infty ds\int_{-\infty}^\infty  dt \frac 1{t-is} 
e^{-[(s+iz_1)^2 +(t-z_2)^2]/16a^2}.
\ee
Using polar coordinates we obtain
\be
Z(z_1,z_2) =\frac{z_2-z_1}{16\pi a^2} \int dr d\phi e^{i\phi+i\theta}
e^{-[r^2 +z_2^2-z_1^2  -2r\sqrt{z_2^2-z_1^2}\cos\phi]/16a^2},
\ee
with 
\be
e^{i\theta} = \frac{z_2+z_1}{\sqrt{z_2^2-z_1^2}}.
\ee
 The integral over $\phi$ is a modified Bessel function
\be
Z(z_1,z_2) = \frac{(z_2-z_1)e^{i\theta}}{8a^2} \int dr I_1(r\sqrt{z_2^2-z_1^2}/8a^2)
e^{-[r^2 +z^2_2-z_1^2 ]/16a^2}.
\ee
Using that
\be  \int_0^\infty dx e^{-\alpha x^2} I_1(\beta x)
      = \frac{1}{\beta}(e^{\beta^2/4\alpha}-1)
\ee
we obtain
\be
Z(z_1,z_2) = -e^{-[z^2_2-z_1^2] /16a^2}+1,
\ee
which disagrees with Eq. (\ref{znorm}).
The missing contributions are the Efetov-Wegner terms which arise due to nilpotent
terms at the singularity of the measure. Below we will evaluate these terms by
regularizing the singularity.

We regularize the integral (\ref{zgdef}) over the matrix elements 
of $\sigma$ by 
introducing the factor
\be 
\theta(\sqrt{a^2 +b^2} - \epsilon).
\ee
Writing out the decomposition (\ref{diag}) we obtain
\be
a&=& s -i\alpha\beta (is-t),\nn \\
b& =& t + \alpha \beta (is-t)
\ee
so that
\be
\sqrt{a^2+b^2} = \sqrt{s^2+t^2} -\frac{\alpha\beta(is-t)^2}{\sqrt{s^2+t^2}},
\ee
and 
\be
\theta( \sqrt{a^2+b^2}-\epsilon) = \theta(\sqrt{s^2+t^2} -\epsilon) 
- \delta(\sqrt{s^2+t^2} -\epsilon) \frac{\alpha\beta(is-t)^2}{\sqrt{s^2+t^2}}.
\ee
The measure is given by
\be
d\sigma  = \frac {ds dt d\alpha d\beta}{(t-is)^2}
\ee
resulting in
\be
Z(z_1,z_2)= \frac 1{2\pi} 
\int  \frac {ds dt d\alpha d\beta}{(t-is)^2} \theta((a^2+b^2)^{1/2}-\epsilon)
e^{[(is-z_1)^2 -(t-z_2)^2 -2\alpha\beta(z_2-z_1)(is-t) ]/16a^2}.
\ee
Expanding the nilpotents in the exponent reproduces the result obtained from the
Itzykson-Zuber integral which does not have to be regularized. We thus find
\be
Z(z_1,z_2)= e^{[z_1^2 -z_2^2]/16a^2} + 
 \frac { z_2-z_1}{16\pi a^2} 
\int  \frac {ds dt} {t-is} 
e^{[(is-z_1)^2 -(t-z_2)^2 ]/16a^2} = 1.
\label{total}
\ee
\begin{center}
\begin{figure}
\includegraphics[width=8.5cm,angle=0]{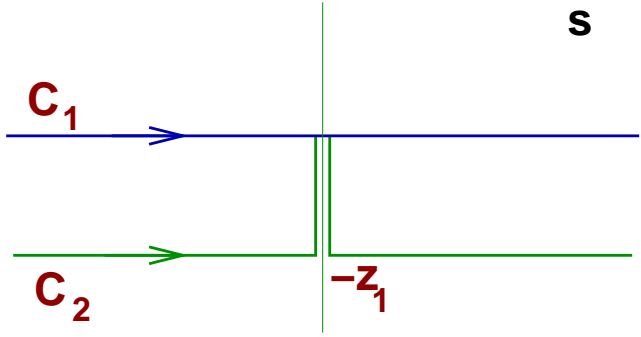}
\caption{\label{fig:cont} Shifting the $s$-integration from $C_1$
to $C_2$ gives an additional contribution from the discontinuity across
the imaginary $s$-axis.}
\end{figure}
\end{center}
This is not the end of the story. Because of a discontinuity in the
integrand we cannot simply shift the integration over $s$ by $-iz_1$. There
is an additional contribution from the discontinuity across the imaginary
$s$-axis. We have that (see Fig. \ref{fig:cont})
\be
\int_{C_1}ds F(s) = \int_{C_2}ds F(s)
\ee
with
\be
\int_{C_2} ds F(s) = \int_{-\infty}^\infty dsF(s-iz_1) + i\int_0^{z_1}dy F(-iz_1+iy-\epsilon)- i\int_0^{z_1}dy F(-iz_1+iy+\epsilon) .\nn\\
\ee
Applying this to the integral in Eq. (\ref{total}) we obtain for the
contribution of the vertical part of the integration contour 
\be
I_\Delta &\equiv&   \frac { z_2-z_1}{16\pi a^2} 
\int dy dt 2\pi \delta(t-z_1+y)   
e^{[y^2 -(t-z_2)^2  ]/16a^2}\nn\\
&=&\frac { z_2-z_1}{8 a^2} 
\int_0^{z_1} dy    
e^{[2y(z_1-z_2) -(z_1-z_2)^2  ]/16a^2},\nn\\
&=&e^{-(z_1-z_2)^2/16a^2} -e^{[z_1^2-z_2^2]/16a^2}.
\ee
The second term cancels against the Efetov-Wegner term. 

The same derivation can be applied to the calculation of the quenched
resolvent. The conclusion is that if we shift the $z$ and $z'$ dependence
from the exponent to the $1/(t-is)$ factor, the Efetov-Wegner term is
of the form $\exp(-(z-z')^2/16a^2)$ which does not contribute to the
quenched resolvent.

\section{Diffusive Partition Function}

\label{sec:partition}
The $N_f$-flavor fermionic partition function was derived in Section
\ref{sec:universality} from the chiral Lagrangian. 
 Including the normalization the $N_f$ flavor partition
function in the sector with index $\nu$ is given by
\be
Z_{ N_f}^\nu(m, {\cal Z};a) &=& \frac{e^{-2N_fa^2}}{(16\pi a^2)^{N_f/2}} \int \prod d s_k
 \Delta(\{is_k\}) \Delta(\{is_k/8a^2\})
 \\ && \hspace*{-2cm}\times
e^{-[(s_1+iz_1)^2+\cdots (s_{N_f}+iz_{N_f})^2 ]/16a^2}{\prod_k(m-is_k)^\nu}
\tilde Z_{ N_f}^\nu(({m^2+s_1^2})^{1/2}, \cdots, ({m^2+s_{N_f}^2})^{1/2}; a=0).\nn
\ee
The normalization factor is such that we recover an identity for $a \to 0$.

For $N_f = 1$ we find
\be
 Z_{N_f =1}^\nu(m,z;a) = \frac {e^{-2a^2}}{\sqrt{16\pi a^2}} \int_{-\infty}^\infty ds e^{-(s+iz)^2/(16a^2)}
(is-m)^\nu \frac{I_\nu(({s^2+m^2})^{1/2})}{(s^2+m^2)^{\nu/2}}.
\label{zone}
\ee
Using the identity
\be
\left ( \frac{is-m}{is+m} \right )^{\nu/2} I_\nu (({s^2+m^2})^{1/2})
= \int_{-\pi}^\pi \frac{d \theta}{2 \pi}e^{i\nu \theta} e^{-m \sin \theta +  s \cos \theta}
\ee
the integral over $s$ becomes a simple Gaussian integral, and the one-flavor
partition function
can be rewritten as
\be
 Z_{N_f =1}^\nu(m,z;a)
&=&
e^{-2a^2} \int_{-\pi}^\pi \frac{d\theta}{2\pi}  e^{i\nu\theta}\ e^{-m\sin\theta-iz\cos \theta + 4a^2
  \sin^2\theta}
\ee
This is indeed the expression for the  one flavor partition derived in
\cite{DSV}.

For $N_f=2$ the normalization factor is given by ${\cal N} =                                                                                                  
1/(\pi(16a^2)^2)$, so that the two flavor partition function reduces to
\be
Z^\nu_{N_f=2}(m_1, m_2;a) &= &\frac {e^{4a^2}}{\pi 8a^2}
\int_{-\infty}^\infty\int_{-\infty}^\infty ds_1 ds_2
\frac{(is_1-is_2)}{m_1-m_2} e^{-\frac 1{16a^2}[(s_1+im_1)^2+ (s_2+im_2)^2 ]}(is_1)^\nu(is_2)^\nu \nn \\
&& \hspace{4cm} \times \tilde Z^\nu_2(({(is_1)^2})^{1/2},({(is_2)^2})^{1/2};a=0),
\label{ztwo}
\ee
where
\be
\tilde Z^\nu_2(x_1,x_2) = \frac 2{x_1^\nu x_2^\nu (x_2^2-x_1^2)} \det \left |
\begin{array}{cc} I_\nu(x_1) & x_1 I_{\nu+1}(x_1)
\\ I_\nu(x_2)& x_2 I_{\nu+1}(x_2)
\end{array}  \right | .
\ee
It is also instructive to work out the partition function for $N_f =-1$.
Using the general expression (\ref{collect}) we obtain
\be
Z^\nu_{N_f=-1}(m,z;a) = \frac {e^{2a^2}}{\sqrt{16\pi a^2}}\int_{-\infty} ^\infty dt e^{-(t-z)^2/16a^2} \frac {(m^2-t^2)^{\nu/2}}
{(t-m)^\nu} (-1)^\nu K_\nu(({m^2-t^2})^{1/2}).
\ee
Using the identity
\be
2\left (\frac{t-i\epsilon +m}{t-i\epsilon -m} \right )^{\nu/2}
K_\nu(({m^2-(t-i\epsilon)^2})^{1/2})
=\int_{-\infty}^\infty ds \ e^{-\nu s} e^{-im\sinh s -i(t-i\epsilon) \cosh s}
\ee
we obtain after performing the Gaussian integration over $t$ and shifting
the $s$-integration by $\pi i$
\be
Z^\nu_{N_f=-1}(m,z;a) &=&
e^{2a^2}\int_{-\infty}^\infty ds \ e^{-\nu s} e^{im \sinh s +iz\cosh s-4a^2\cosh^2s} \nn \\
&=&\int_{-\infty}^\infty ds \ e^{-\nu s} e^{-im \sinh                                                                                                         
  s -iz\cosh s-2a^2\cosh(2s)} .
\ee
which agrees with
the bosonic part of the result obtained in \cite{DSV}.
 


\begin{thebibliography}{99}


\bibitem{BC}
  T.~Banks, A.~Casher,
  Nucl.\ Phys.\  {\bf B169}, 103 (1980).
 

\bibitem{Heller}
  K.~M.~Bitar, U.~M.~Heller and R.~Narayanan,
  Phys.\ Lett.\  B {\bf 418}, 167 (1998).
[arXiv:hep-th/9710052].

\bibitem{Aokiclassic}
S.~Aoki,
  Phys.\ Rev.\  D {\bf 30} (1984) 2653.



\bibitem{Luscher}
  L.~Del Debbio, L.~Giusti, M.~L\"uscher, R.~Petronzio and N.~Tantalo,
  JHEP {\bf 0602}, 011 (2006)
  [hep-lat/0512021];
  JHEP {\bf 0702}, 056 (2007)
  [hep-lat/0610059].

\bibitem{DSV}
  P.~H.~Damgaard, K.~Splittorff and J.~J.~M.~Verbaarschot,
  Phys.\ Rev.\ Lett.\  {\bf 105}, 162002 (2010).
  [arXiv:1001.2937 [hep-th]].


\bibitem{SV}
  E.~V.~Shuryak and J.~J.~M.~Verbaarschot,
  Nucl.\ Phys.\  A {\bf 560}, 306 (1993)
  [hep-th/9212088].

\bibitem{RMT}
  J.~J.~M.~Verbaarschot and I.~Zahed,
  Phys.\ Rev.\ Lett.\  {\bf 70}, 3852 (1993)
  [hep-th/9303012];
  J.~J.~M.~Verbaarschot,
  Phys.\ Rev.\ Lett.\  {\bf 72}, 2531 (1994)
  [hep-th/9401059].
  Nucl.\ Phys.\  B {\bf 426}, 559 (1994)
  [hep-th/9401092].
  A.~D.~Jackson, M.~K.~Sener and J.~J.~M.~Verbaarschot,
  Phys.\ Lett.\  B {\bf 387}, 355 (1996)
  [hep-th/9605183].
  M.~K.~Sener and J.~J.~M.~Verbaarschot,
  Phys.\ Rev.\ Lett.\  {\bf 81}, 248 (1998)
  [hep-th/9801042].


\bibitem{RMT1}
  A.~M.~Halasz and J.~J.~M.~Verbaarschot,
  Phys.\ Rev.\  D {\bf 52}, 2563 (1995)
  [hep-th/9502096].

\bibitem{RMT2}
  G.~Akemann, P.~H.~Damgaard, U.~Magnea and S.~Nishigaki,
  Nucl.\ Phys.\  B {\bf 487}, 721 (1997)
  [hep-th/9609174].
  P.~H.~Damgaard and S.~M.~Nishigaki,
  Nucl.\ Phys.\  B {\bf 518}, 495 (1998)
  [hep-th/9711023].
  P.~H.~Damgaard,
  Phys.\ Lett.\  B {\bf 424}, 322 (1998)
  [arXiv:hep-th/9711110].
  G.~Akemann and P.~H.~Damgaard,
  Nucl.\ Phys.\  B {\bf 528}, 411 (1998)
  [arXiv:hep-th/9801133].


\bibitem{ADSVprd}
  G.~Akemann, P.~H.~Damgaard, K.~Splittorff, J.~J.~M.~Verbaarschot,
  Phys. Rev. D 83, 085014 (2011) [arXiv:1012.0752 [hep-lat]].


\bibitem{ADSVNf1}
  G.~Akemann, P.~H.~Damgaard, K.~Splittorff, J.~J.~M.~Verbaarschot,
  PoS {\bf LATTICE2010}, 079 (2010).
  [arXiv:1011.5121 [hep-lat]].

\bibitem{GuhrJMath}
 T.~Guhr, J. Math. Phys. {\bf 32}, 336 (1991).


\bibitem{Alfaro-1994}
  J.~Alfaro, R.~Medina, L.~F.~Urrutia,
  J.\ Math.\ Phys.\  {\bf 36}, 3085-3093 (1995).
  [hep-th/9412012].


\bibitem{GuhrAnn}
  T.~Guhr,
  Annals Phys.\  {\bf 250}, 145-192 (1996).
  
\bibitem{GuhrComm}
  T.~Guhr,
  Commun.\ Math.\ Phys.\  {\bf 176}, 555-576 (1996).
  


\bibitem{SharpeSingleton}
  S.~R.~Sharpe and R.~L.~Singleton,
  Phys.\ Rev.\  D {\bf 58}, 074501 (1998)
  [hep-lat/9804028].

\bibitem{RS}
G.~Rupak and N.~Shoresh,
Phys.\ Rev. \ {\bf 66}, 054503 (2002), [arXiv:hep-lat/0201019].


\bibitem{BRS}
O.~B\"ar, G.~Rupak and N.~Shoresh,
Phys.\ Rev. \ D {\bf 70}, 034508 (2004), [arXiv:hep-lat/0306021].

\bibitem{Aoki-spec}
  S.~Aoki,
  Phys.\ Rev.\  D {\bf 68}, 054508 (2003)
  [arXiv:hep-lat/0306027].


\bibitem{Aoki-pion-mass}
  S.~Aoki and O.~B\"ar,
  Phys.\ Rev.\  D {\bf 70}, 116011 (2004)
  [arXiv:hep-lat/0409006].


 \bibitem{bar-08}
 O.~Bar, S.~Necco, S.~Schaefer,
 JHEP {\bf 0903}, 006 (2009).
 [arXiv:0812.2403 [hep-lat]].


 \bibitem{shindler-09}
 A.~Shindler,
 Phys.\ Lett.\  {\bf B672}, 82-88 (2009).
 [arXiv:0812.2251 [hep-lat]].



\bibitem{Golterman}
  M.~Golterman,
  arXiv:0912.4042.



\bibitem{sharpe-nara}
  S.~R.~Sharpe,
  arXiv:hep-lat/0607016.


\bibitem{sharpe}
  S.~R.~Sharpe,
  Phys.\ Rev.\  D {\bf 74}, 014512 (2006)
  [arXiv:hep-lat/0606002].

\bibitem{NS}
  S.~Necco, A.~Shindler,
  [arXiv:1101.1778 [hep-lat]].


\bibitem{DOTV}
  P.~H.~Damgaard, J.~C.~Osborn, D.~Toublan and J.~J.~M.~Verbaarschot,
  Nucl.\ Phys.\  B {\bf 547}, 305 (1999)
  [hep-th/9811212].


\bibitem{splitchem}
 K.~Splittorff, J.~J.~M.~Verbaarschot,
 Nucl.\ Phys.\  {\bf B683}, 467-507 (2004).
 [hep-th/0310271];
 K.~Splittorff, J.~J.~M.~Verbaarschot,
 Nucl.\ Phys.\  {\bf B757}, 259-279 (2006).
 [hep-th/0605143];
 K.~Splittorff, J.~J.~M.~Verbaarschot, M.~R.~Zirnbauer,
 Nucl.\ Phys.\  {\bf B803}, 381-404 (2008).
 [arXiv:0802.2660 [hep-th]].

\bibitem{Efetov} K.~B.~Efetov, {\it Supersymmetry in Disorder and Chaos}, Cambridge University Press, Cambridge, 1997.

\bibitem{Guhr:1996vx}
  T.~Guhr, T.~Wettig,
  J.\ Math.\ Phys.\  {\bf 37}, 6395-6413 (1996).
  [hep-th/9605110].


 
\bibitem{SplitVerb1}
  K.~Splittorff and J.~J.~M.~Verbaarschot,
  Phys.\ Rev.\ Lett.\  {\bf 90}, 041601 (2003)
  [cond-mat/0209594].
 

\bibitem{FA}
  Y.~V.~Fyodorov and G.~Akemann,
  JETP Lett.\  {\bf 77}, 438 (2003)
  [Pisma Zh.\ Eksp.\ Teor.\ Fiz.\  {\bf 77}, 513 (2003)]
  [cond-mat/0210647].


\bibitem{Itoh}
  S.~Itoh, Y.~Iwasaki and T.~Yoshie,
  Phys.\ Rev.\  D {\bf 36}, 527 (1987).

\bibitem{mario}
  M.~Kieburg,
    [arXiv:1011.0836 [math-ph]].

\end{thebibliography}
\end{document}